\documentclass{aastex63}

\submitjournal{Frontiers}

\graphicspath{{./}{figures/Results Plots/}}
\usepackage{url}
\usepackage{amssymb}
\usepackage{dsfont}
\usepackage{amsmath}
\usepackage{float}
\usepackage{longtable}

\begin{document}

\title{Using Multivariate Imputation by Chained Equations to Predict Redshifts of Active Galactic Nuclei}

\author{Spencer James Gibson}
\affiliation{Carnegie Mellon University, USA}

\author{Aditya Narendra}
\affiliation{Jagiellonian University, Poland}

\correspondingauthor{Maria Giovanna Dainotti}
\email{maria.dainotti@nao.ac.jp}

\author{Maria Giovanna Dainotti}
\affiliation{National Astronomical Observatory of Japan, Mitaka}
\affiliation{Space Science Institute, 4750 Walnut St, Suite 205, Boulder,CO,80301,USA}

\author{Malgorzata Bogdan}
\affiliation{Department of Mathematics, University of Wroclaw, Poland}
\affiliation{Department of Statistics, Lund University, Sweden}

\author{Agnieszka Pollo}
\affiliation{Astronomical Observatory of Jagiellonian University, Krakow}
\affiliation{National Centre for Nuclear Research, Warsaw}

\author{Artem Poliszczuk}
\affiliation{National Centre for Nuclear Research, Warsaw}

\author{Enrico Rinaldi}
    \affiliation{Physics Department, University of Michigan, Ann Arbor, MI 48109, USA}
    \affiliation{Theoretical Quantum Physics Laboratory, RIKEN, 2-1 Hirosawa, Wako, Saitama, 351-0198, Japan}
    \affiliation{Interdisciplinary Theoretical \& Mathematical Science Program, RIKEN (iTHEMS), 2-1 Hirosawa, Wako, Saitama, 351-0198, Japan}
    
\author{Ioannis Liodakis}
\affiliation{Finnish Centre for Astronomy with ESO (FINCA), University of Turku, Finland}

\begin{abstract}
    Redshift measurement of active galactic nuclei (AGNs) remains a time-consuming and challenging task, as it requires follow up spectroscopic observations and detailed analysis.
    Hence, there exists an urgent requirement for alternative redshift estimation techniques. 
    The use of machine learning (ML) for this purpose has been growing over the last few years, primarily due to the availability of large-scale galactic surveys.
    However, due to observational errors, a significant fraction of these data sets often have missing entries, rendering that fraction unusable for ML regression applications.
    In this study, we demonstrate the performance of an imputation technique called Multivariate Imputation by Chained Equations (MICE), which rectifies the issue of missing data entries by imputing them using the available information in the catalog. We use the Fermi-LAT Fourth Data Release Catalog (4LAC) and impute 24\% of the catalog. Subsequently, we follow the methodology described in \cite{dainotti2021predicting} and create an ML model for estimating the redshift of 4LAC AGNs. 
    We present results which highlight positive impact of MICE imputation technique on the machine learning models performance and obtained redshift estimation accuracy.
\end{abstract}

\section{Introduction}
Spectroscopic redshift measurement of Active Galactic Nuclei (AGNs) is a highly time-consuming operation and is a strong limiting factor for a large-scale extragalactic surveys.
Hence, there is a pressing requirement for alternative redshift estimation techniques that provide reasonably good results \citep{salvato2019NatAs_photoz_rev}. 
In current cosmological studies, such alternative redshift estimates, referred to as photometric redshifts, play a key role in our understanding of the Extragalactic Background Light (EBL) origins \citep{Wakely2008}\footnote{http://tevcat.uchicago.edu/}, magnetic field structure in the intergalactic medium \citep{2020AAS...23540506M,2013MNRAS.432.3485V,2018Sci...362.1031F}
and help in determining the bounds on various cosmological parameters \citep{2019ApJ...885..137D,petrosian1976surface,singal2013cosmological,singal2012flux,singal2014gamma,singal2015determination,singal2013flat,chiang1995evolution,ackermann2015multiwavelength,singal2013cosmological,ackermann2012gev}.
\newline\indent
One technique that has gained significant momentum is the use of machine learning (ML) to determine the photometric redshift of AGNs \citep{brescia2013photometric,brescia2019photometric,dainotti2021predicting,nakoneczny2020photometric,jones2017analysis,cavuoti2014photometric,fotopoulou2018cpz,logan2020unsupervised,yang2017quasar,zhang2019machine,curran2020qso,  nakoneczny2020photometric,pasquet2018deep,jones2017analysis}. 
Large AGN data sets derived from all-sky surveys like the Wide-field Infrared Survey Explorer (WISE) \citep{Brescia2019,ilbert2008cosmos,hildebrandt2010phat,brescia2013photometric,2010,DIsanto2018} and Sloan Digital Sky Survey (SDSS) \citep{aihara2011eighth} have played a significant role in the proliferation of ML approaches. 
However, the quality of the results from an ML approach depends significantly on the size and quality of the training data: the data on which the ML models learn the underlying relationship to predict the redshift.
Unfortunately, almost all of these large data sets suffer from the issue of missing entries, which can lead to a considerable portion of the data being discarded.

This is especially problematic in catalogs of smaller size, such as in the case of gamma-ray loud AGNs.

\indent
Using the Fermi Fourth Data Release Catalog's (4LAC) gamma-ray loud AGNs \citep{2020ApJ...892..105A,2020ApJS..247...33A}, \cite{dainotti2021predicting} demonstrated that ML methods lead to promising results, with a 71\% correlation between the predicted and observed redshifts.
However, in that study, the training set consists of only 730 AGNs, and a majority of the data (50\%) are discarded due to missing entries.
{\bf More specifically, we have several reasons why the sources are missing also in relation to the variables we consider. Regarding the missing values of the Gaia magnitudes: this could be either because the sources are too faint and thus they undergo the so called Malmquist bias effect (only the brightest sources are visible at high-z) or the coordinates are not accurate enough and the cross-matching is failing to produce a counterpart (the latter is not that likely, the former is much more likely)

Regarding the variables observed in $\gamma$-rays: here the source is detected, but it is faint in gamma-rays and again we have the Malmquist bias effect in relation to the detector threshold of Fermi-LAT and/or it does not appear variable and/or the spectral fitting fails to produce values, hence the missing values.

Regarding the multi-wavelength estimates ($\nu$, $\nu f_\nu$): these depend on the availability of multi-wavelength data from radio to X-rays. If sufficient data exists then a value can be estimated, so the missing values are most likely sources that have not been observed by telescopes.  In other words, this does not mean that the sources are necessarily faint, they could be bright, but just no telescope performed follow-up observations.

There is also the possibility to explain the missing values because of the relativistic effects that dominate blazar emission. The relativistic effects, quantified by a parameter called the Doppler factor, boost the observed flux across all frequencies, but also shorten the timescales making sources appear more variable. It has been shown that sources detected in $\gamma$-rays have higher Doppler factors and are more variable \citep{Liodakis2017,Liodakis2018}. This would suggest that sources observed more off-axis, i.e., lower Doppler factor, would have a lower $\gamma$-ray flux and appear less variable. Therefore introduce more missing values as we have discussed above.
}

In this study, we address this issue of missing entries using an imputation technique called Multivariate Imputation by Chained Equations (MICE) \citep{van2011mice}.
This technique was also recently used by \cite{luken2021missing} for redshift estimation of Radio-loud AGNs.
\newline\indent
\cite{luken2021missing} test multiple imputation techniques, MICE included, to determine the
best tool for reliably imputing missing values. 
Their study considers the redshift estimation of radio-loud
galaxies present in the Australia Telescope Large Area Survey (ATLAS). 
However, in contrast to our approach where we impute actual missing information in the catalog, they manually set specific percentages of their data as missing and test how effective various imputation techniques are.
Their results demonstrate distinctly that MICE is the best imputation technique, leading to the least root mean square error (RMSE) and outlier percentages for the regression algorithms they have tested. 
\newline\indent
In our study, we are using the updated 4LAC catalog, and using MICE imputations to fill in missing entries, we achieve a training data set which is 98\% larger than the one used in \cite{dainotti2021predicting}.
We achieve results on this more extensive training set that are comparable to \cite{dainotti2021predicting} while attaining higher correlations.
Furthermore, we are using additional ML algorithms in the SuperLearner ensemble technique, as compared to \cite{dainotti2021predicting}.
\newline\indent

Sec.~\ref{sec:Sample} discusses the specifics of the extended 4LAC data set: how we create the training set, which predictors are used and which outliers are removed. 
In Sec.~\ref{sec:Methodology} we discuss the MICE imputation technique, the SuperLearner ensemble with a brief description of the six algorithms used in this analysis, followed by the different feature engineering techniques implemented. 
Finally, we present the results in Sec.~\ref{sec:Results}, followed by the discussion and conclusions in Sec.~\ref{sec:D&C}.

\section{Sample}\label{sec:Sample}
\begin{figure}[H]
\centering
\includegraphics[width=0.75\textwidth]{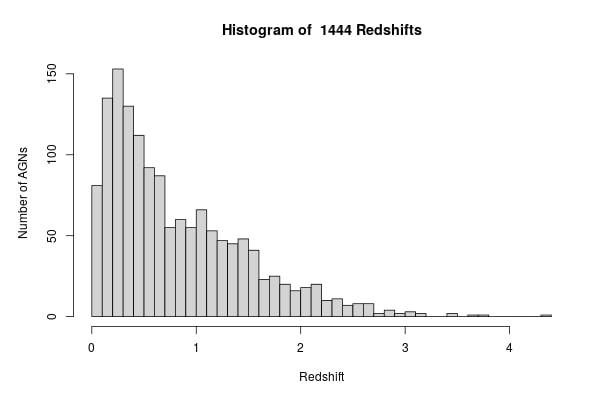}
\caption{Redshift distribution of the training set. }
\label{fig:red1}
\end{figure}

This study uses the Fermi Fourth Data Release Catalog (4LAC), containing 3511 gamma-ray loud AGNs, 1764 of which have a measured spectroscopic redshift.
Two categories of AGNs dominate the 4LAC catalog, BL Lacertae (BLL) objects and Flat Spectrum Radio Quasars (FSRQ). 
To keep the analysis consistent with \cite{dainotti2021predicting}, we remove all the non-BLL and non-FSRQ AGNs.
\newline\indent
These AGNs have 13 measured properties in the 4LAC catalog; however, we only use 11 and a categorical variable that distinguishes BLLs and FSRQs. 
The two omitted properties in the analysis are $Highest\_Energy$ and $Fractional\_Variability$ because 42.5\% of the entries are missing, and there is insufficient information to impute them reliably.
We consider imputation of predictors which have missing entries in less than 18\% of the data.
The remaining 11 properties and the categorical variables are $Gaia\_G\_Magnitude$, $Variability\_Index$, $Flux$, $Energy\_Flux$, $PL\_Index$, $\nu f\nu$, $LP\_Index$, $Significance$, $Pivot\_Energy$, $\nu$, and $LP\_\beta$ and $LabelNo$, serve as the predictors for the redshift in the machine learning models and are defined in \cite{dainotti2021predicting} and \cite{2020ApJ...892..105A}. 
However, some of these properties are not used as they appear in the 4LAC, since they span several orders of magnitude. The properties $Flux, Energy\_Flux, Significance, Variability\_Index, \nu, \nu$f$\nu $, and $ Pivot\_Energy$ are used in their base-10 logarithmic form.
In the categorical variable $LabelNo$ we assign the values 2 and 3 to BLLs and FSRQs, respectively. 
We are not training the ML models to predict the redshift directly. 
Instead, we train the models to predict $1/(z+1)$, where $z$ is the redshift. 
Such a transformation of the target variable is crucial as it helps improve the model's performance. 
In addition, $1/(z+1)$ is known as the scale factor and has a more substantial cosmological significance than redshift itself.
We remove AGNs with an $LP\_\beta$ $<$ 0.7, $LP\_Index$ $>$ 1, and $LogFlux$ $>$ -10.5, as they are outliers of their respective distributions.
These steps lead us to a final data sample of 1897 AGNs, out of which 1444 AGNs have a measured redshift (see Fig. \ref{fig:red1}). 
These AGNs form the training sample, while the remaining 453 AGNs, which do not have a measured redshift, form the generalization sample.

\section{Methodology}\label{sec:Methodology}
Here we present the various techniques implemented in the study, definitions of the statistical metrics used, and a comprehensive step-by-step description of our procedure to obtain the results.
We use the following metrics to measure the performance of our ML model:
\begin{itemize}
    \item Bias: Mean of the difference between the observed and predicted values.
    \item $\sigma_{NMAD}$: Normalized median absolute deviation between the predicted and observed measurements.
    \item $r$: Pearson correlation coefficient between the predicted and observed measurements.
    \item Root Mean Square Error (RMSE) between the predicted and observed redshift
    \item Standard Deviation $\sigma$ between the predicted and observed redshift
\end{itemize}
We present these metrics for both $\Delta z_{norm}$ and $\Delta z$, which are defined as:
\begin{equation}
    \Delta z = z_{observed} - z_{predicted} 
\end{equation}

\begin{equation}
 \Delta z_{norm} = \frac{\Delta z}{1 + z_{observed}}
\end{equation}

We also quote the catastrophic outlier percentage, defined as the percentage of predictions that lies beyond the 2$\sigma$ error.
The metrics presented in this study are the same as in \cite{dainotti2021predicting}, allowing for easy comparison.

\subsection{Procedure}\label{sec:procedure}
Here we provide a walk-through of how the final results are obtained.
First, we remove all the non-BLL and non-FSRQ AGNs from the 4LAC data set, in addition to outliers, and end up with 1897 AGNs for the total set. 
Then, we impute the missing entries using MICE (see Sec. \ref{sec:MICE}).
Having obtained a complete data set, we split it into the training and the generalization sets, depending on whether the AGNs have or do not have a measured redshift value.
We aim to train an ensemble model that is the least complex and best suited to the data at hand.
For this purpose, we need to test many different algorithms with ten-fold cross-validation (10fCV). Cross-validation is a resampling procedure that uses different portions of the data, in this case 10, to train and test a model, and find out which algorithm performs the best in terms of the previously defined metrics.
However, since there is inherent randomness in how the folds are created during 10fCV, we perform 10fCV one hundred times and average the results to derandomize and stabilize them.
This repeated k-fold cross-validation technique is standard in evaluating ML models.
In each of the one hundred iterations of 10fCV, we train a SuperLearner model (see Sec. \ref{sec:SL}) on the training set using the twelve algorithms shown in Fig. \ref{fig:all_algo}.
Finally, averaging over the one hundred iterations, we obtained the coefficients and risk measurements associated with each SuperLearner ensemble model, as well as the individual algorithms.
Following the previous step, we pick six algorithms that have coefficients greater than 0.05 (see Sec. \ref{sec:ML} for information about these algorithms). 
\newline\indent
With the six best ML algorithms, we create an ensemble with SuperLearner and perform the 10fCV one hundred times once more.
The final cross-validated results are again an average of these one hundred iterations. 
\indent
Next, we proceed to show the results obtained without the repeated cross-validation procedure.
For this, we simply select a fixed validation set by choosing the last 111 AGNs from the 1444 AGNs of the previously used training data. 
Now, with the new training set of 1333 AGNs, we train a SuperLearner model, with the algorithms being the same as in the cross-validation step, and we predict the redshift of the validation set.
We then calculate the same statistical metrics for these results as we did for the cross-validated results. 
The results on this fixed validation set provide a representative of the performance of the SuperLearner model, which we have explored in more details (and in a more computationally expensive way) during the repeated cross-validation procedure.

\subsection{MICE}\label{sec:MICE}
Multivariate Imputation by Chained Equations (MICE) is a method for imputing missing values for multivariate data \citep{van2011mice,luken2021missing}. 
The multivariate in MICE highlights its use of multiple variables to impute missing values.
The MICE algorithm works under the assumption that the data are missing at random (MAR). 
MAR was first detailed in the paper \cite{rubin1976inference}.
It implies that errors in the system or with users cause the missing entries and not intrinsic features of the object being measured.
Furthermore, MAR implies the possibility that the missing entries can be inferred by the other variables present in the data \citep{schafer2002missing}. 
Indeed, this is a strong assumption, and it is our first step to deal with missing data. However, we know that selection biases play an important role for the flux detection. 
Although this problem is mitigated for the gamma-ray sources, for the G-band magnitude, one can argue that, e.g., BL Lacs are systematically fainter than FSRQs and below the Gaia limiting magnitude.
A more in-depth analysis to take this problem into account is worthwhile, but this is beyond the scope of the current paper.

\indent
With this assumption, MICE attempts to fill in the absent entries using the complete variables in the data set iteratively.  
We impute the missing variables 20 times with each iteration of MICE consisting of multiple steps.
General practice is to perform the imputation ten times as in \cite{luken2021missing} and \cite{van2011mice}, but we perform it twenty times to stabilize the imputation.

Here, we use the method ``midastouch'' - a predictive mean matching (PMM) method \citep{little2019statistical}.
It works by initializing a feature's missing entries with its mean and then estimating them by training a model using the rest of the complete data. 
For each prediction, a probability is assigned based on its distance from the value imputed for the desired entry. 
The missing entry is imputed by randomly drawing from the observed values of the respective predictor, weighted according to the probability defined previously.

The process is repeated for each missing entry until all have been refitted. 
This new complete table is used as a basis for the next iteration of MICE, where the same process is repeated until the sequence of table converges or a set number of iterations is achieved.

\indent

\subsection{SuperLearner}\label{sec:SL}
SuperLearner \citep{van2007super} is an algorithm that constructs an ensemble of ML models predictions using a cross-validated metric and a set of normalized coefficients. By default Superlearner uses a ten-fold cross-validation procedure.
It outputs a combination of user-provided ML models such that the RMSE of the final prediction is minimized by default \citep{polley2010super} (or any other user-defined metric defining the expected risk of the task at hand).
In our setup, SuperLearner achieves this using 10fCV, where the training data is divided into ten equal portions or folds, the models are trained on nine folds, and the 10th fold is used as a test set.
The models predict the target variable of the test set, and based on the RMSE of their predictions, SuperLearner assigns a coefficient.
If an algorithm has a lower RMSE in 10fCV, it will be assigned a higher coefficient.
Finally, it creates the ensemble as a linear combination of the constituent models multiplied by their respective coefficients.
Note that this 10fCV is an internal procedure of model selection to build the SuperLearner ensemble model, and it is separate from the repeated cross-validation procedure which we described in Sec.~\ref{sec:procedure} and which is used to evaluate the performance and final results.

\subsection{The ML algorithms used in our analysis}\label{sec:ML}
\begin{figure}[H]
    \centering
    \includegraphics[width=\textwidth]{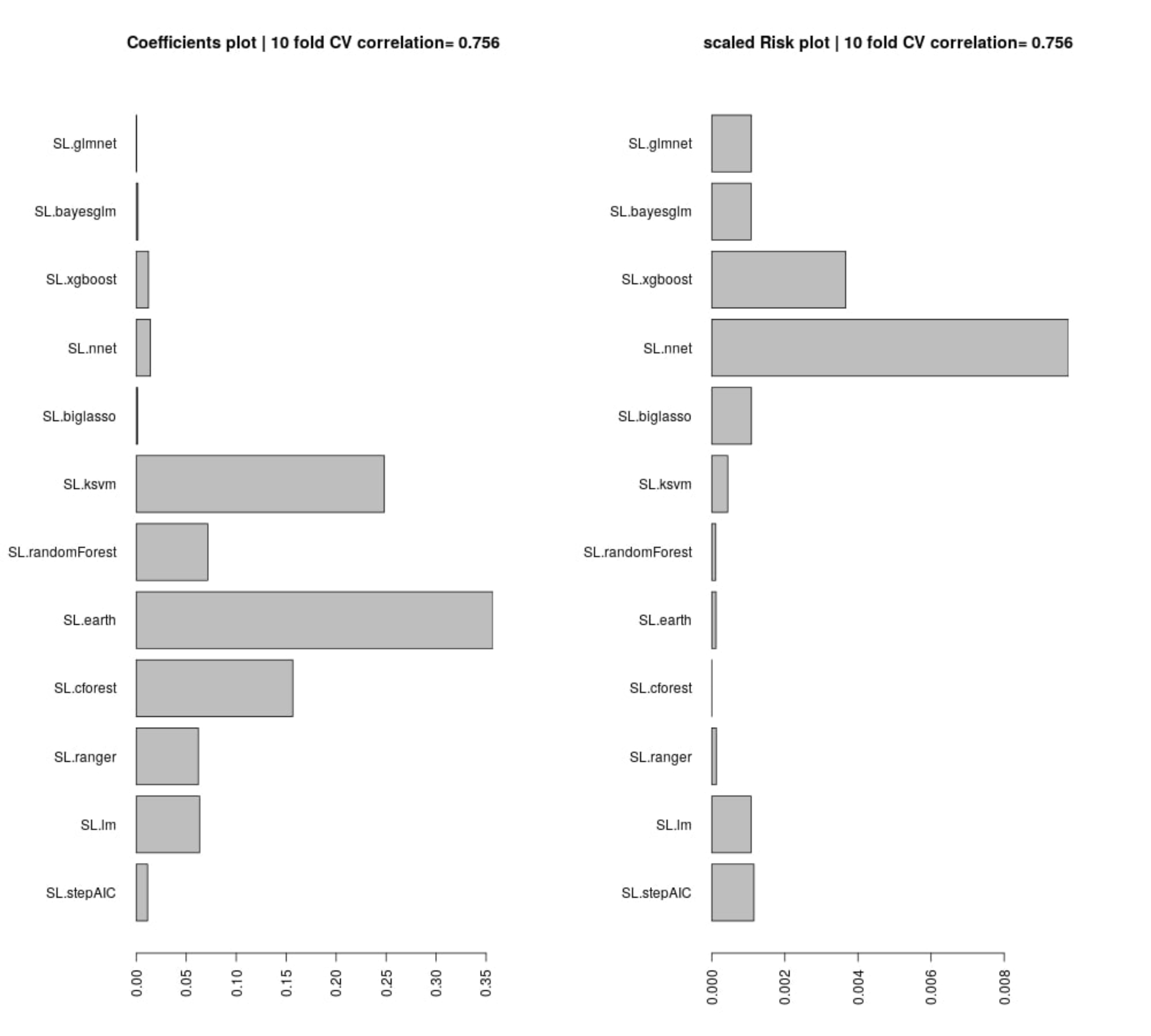}
    \caption{
    Left Panel: The coefficients assigned by SuperLearner to the algorithms tested. We select the algorithms that have a coefficient above 0.05 to be incorporated into our ensemble. 
    Right Panel: The RMSE error (risk) of each of the algorithms, scaled to show the minimum risk algorithm at 0, which is Cforest.
    These values are average over one hundred iterations of 10fCV.
    }
    \label{fig:all_algo}
\end{figure}

Following \cite{dainotti2021predicting} we analyze the coefficients assigned by SuperLearner to twelve ML algorithms, and pick those with a value greater than 0.05.
In Fig. \ref{fig:all_algo}, we show all the ML algorithms tested, and their coefficients. 
We pick the six algorithms above the 0.05 cutoff, which are: Enhanced Adaptive Regression Through Hinge (EARTH), KSVM, Cforest, Ranger, Random Forest, and Linear Model. We provide brief explanations for each of them below.

Enhanced Adaptive Regression Through Hinges (EARTH) is an algorithm that allows for better modeling of predictor interaction and non-linearity in the data compared to the linear model. 
It is based on the Multivariate Adaptive Regression Splines method (MARS) \citep{FriedmanMARS}. 
EARTH works by fitting a sum or product of hinges. 
Hinges are part-wise linear fits of the data that are joined such that the sum-of-squares residual error is minimized with each added term.
\newline\indent
\newline\indent
KSVM is an R implementation of the Support Vector Regression method (SVR). 
Similar to Support Vector Machine (SVM) \citep{cortes1995support}, SVR uses a kernel function to send its inputs to a higher-dimensional space where the data is linearly separable by a hyper-plane.
SVR aims to fit this hyper-plane such that the prediction error is within a pre-specified threshold.
For our purposes, KSVM uses the Gaussian kernel with the default parameters.
\newline\indent
\newline\indent
The Random Forest algorithm \citep{breiman2001random,randomForestsHo} seeks to extend decision trees capabilities by simultaneously generating multiple, independent decision trees. For regression tasks, Random Forest will return the average of the outputs of each of the generated decision trees.
An advantage of Random Forest over decision trees is the reduction in the variance.
However, Random Forest often suffers from low interpretability.
\newline
The Ranger algorithm is similar to Random Forest with the difference of extremely randomized trees (ERTs) \citep{geurts06extremetrees} and quicker implementation. 
\newline
Similar to Random Forest, the Cforest algorithm \citep{hothorn2006unbiased} builds conditional inference trees that perform splits on significance tests instead of information gain.
\newline\indent
We use the ordinary least squares (OLS) linear model found in the SuperLearner package. This model aims to minimize the mean squared error.
\newline\indent
Note that we are using the default hyperparameter settings for all the algorithms.

\subsection{Feature Engineering}\label{sec:featureEngineering}
Feature engineering is a broad term that incorporates two techniques: feature selection and feature creation.
Feature selection is a method where the best predictors of a response variable are chosen from a larger pool of predictors. 
There exist multiple methods to perform feature selection. 
We are using the Least Absolute Selection and Shrinkage Operator (LASSO) method.
Feature selection is an essential part of any ML study as it reduces the dimensionality of the data and minimizes the risk of overfitting. 
Feature creation is a technique where additional features are created from various combinations of existing properties. 
These combinations can be cross-multiplications, higher-order terms, or ratios. 
Feature creation can reveal hidden patterns in the data that ML algorithms might not be able to discern and consequently boost the performance. 

\textbf{In machine learning, some of the methods used  by SuperLearner are linear by nature (BayesGLM, Lasso, elastic-net). Adding quadratic and multiplicative terms allows us to model some types of non-linear relationships.
Interactions among variables are very important and can boost the prediction when used. The phrase "interaction among the variables" means the influence of one variable on the other; however, not in an additive way, but rather in a multiplicative way. In our feature engineering procedure, we build these interactions by cross-products and squares of the initial variables.  It is common that adding O2 predictors aids results since they may contain information not available in the O1 predictors.}

In this study, we create 66 new features, which, as mentioned, are the cross-products and squares of the existing features of the 4LAC catalog.
We denote the existing predictors of the 4LAC catalog as Order-1 (O1) predictors and the new predictors as Order-2 (O2).
Thus, we expand the set of predictors from the initial eleven O1 predictors to a combined seventy-eight O1 and O2 predictors.

For features selection, LASSO \citep{TibshiraniLasso} is used. 
It works by constraining the $\ell^1$ norm of the coefficient vector to be less than or equal to a tuning parameter $\lambda$ while fitting a linear model to the data. 
The predictors that LASSO chooses have a non-zero coefficient for the largest $\lambda$ value with the property that the corresponding prediction error is within one standard deviation of the minimum prediction error \citep{hastieTibs,birnbaum1962foundations,hastie1987generalized,hastie1990generalized,friedman2010regularization}.
This study performs LASSO feature selection on a fold-by-fold basis during external 10fCV.
Optimal features are picked using LASSO for nine of the ten folds, and the predictions on the tenth fold are performed using these selected features. 
This step is iterated such that for every combination of nine folds, an independent set of features is picked. 
This usage of LASSO is in contrast to \cite{dainotti2021predicting}, where the best features are picked for the entire training set. 
Our updated technique ensures that during the 10fCV, LASSO only picks the best predictors based on the training data, and the test set does not affect the models.
This feature selection method is applied to both the O1 and O2 predictor sets.





\section{\textbf{Results}}\label{sec:Results}
\begin{figure}[H]
    \centering
    \includegraphics[width = 0.75\textwidth]{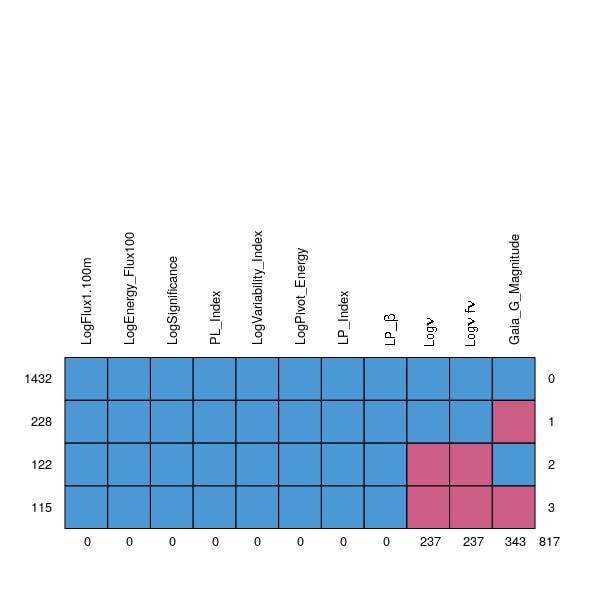}
    \caption{The pattern of the missing data. The blue cells represent complete values, while the pink ones indicate where we have missing data. 
    The first row shows that there are 1432 AGNs without missing values.
    Second row shows that there are 228 data points with $Gaia\_G\_Magnitude$ missing.
    Third row shows that there are 122 data points with $Log\nu$ and $Log\nu f\nu$ missing.
    And finally, the last row shows that there are 115 data points with missing values in $Gaia\_G\_Magnitude$, $Log\nu$ and $Log\nu f\nu$. 
    The columns indicate that there are 237 missing values in $Log\nu$ and $Log\nu f\nu$, and 343 missing values in $Gaia\_G\_Magnitude$. The remaining predictors have no missing entries.
    }
    \label{fig:missing_data}
\end{figure}

\begin{figure}[H]
    \centering
    \includegraphics[width=0.65\textwidth]{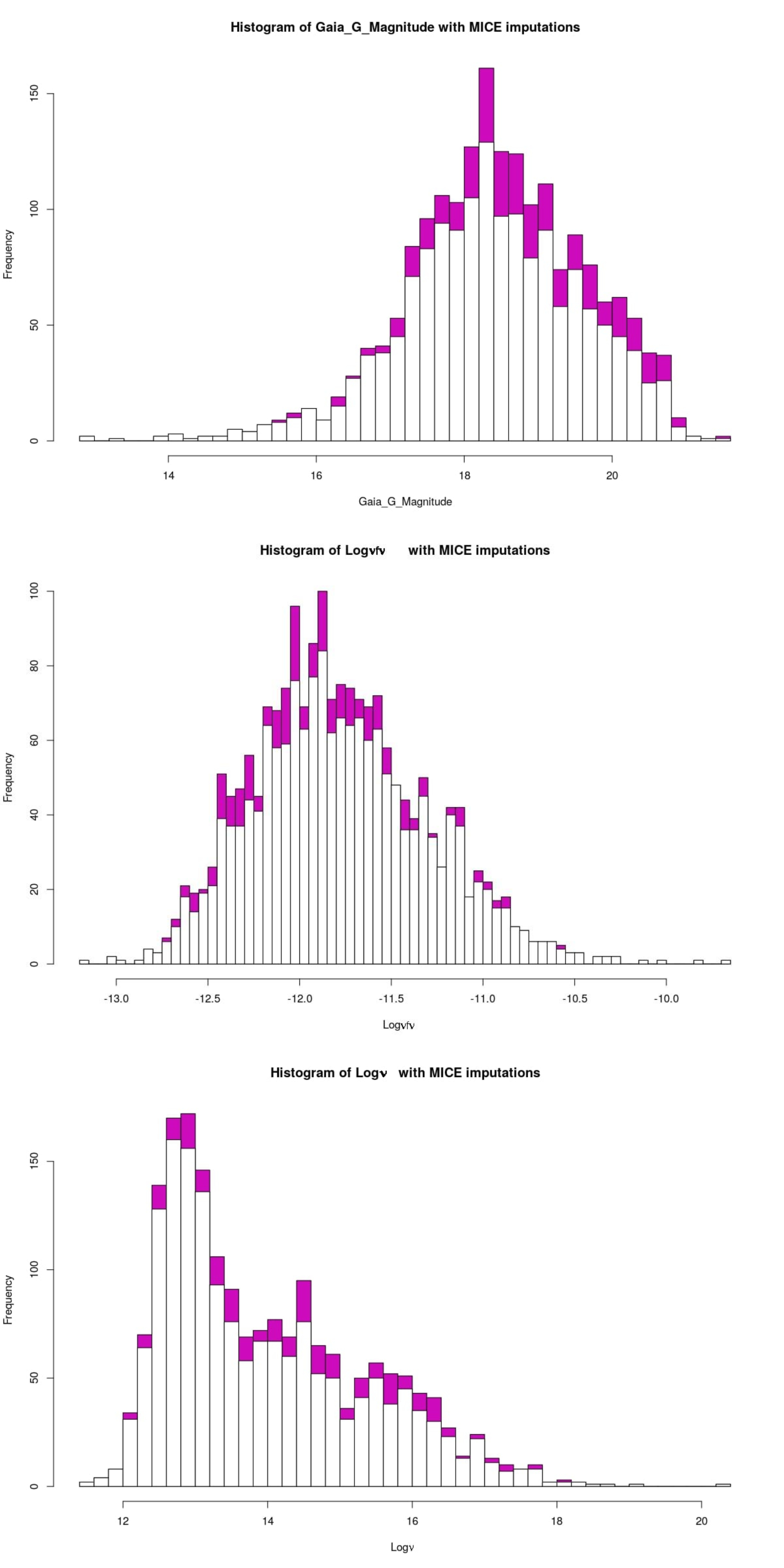}
    \caption{The white bars show the initial distribution of the variables. The magenta bars plotted on top of it are the MICE imputed values. The top plot shows the distribution of $Gaia\_G\_Magnitude$ with and without MICE. The central plot shows this for $Log\nu f\nu$, and the bottom plot shows this for $Log\nu$.}
    \label{fig:mice_imputations}
\end{figure}

The quality of the MICE imputations depends on the information density of the entire data set.
Hence, to ensure the best possible imputations we use all 1897 AGNs which remain after the removal of outliers and non-BLL and non-FSRQ AGNs. 
The pattern of the missing entries in our data set is shown in Fig. \ref{fig:missing_data}, and they are present in only three predictors, namely, $Log\nu$, $Log\nu f\nu$, and $Gaia\_G\_Magnitude$ (see Sec. \ref{sec:Sample}). 
There are 237 AGNs which have missing values in both $Log\nu$ and $Log\nu f\nu$, and 343 AGNs have a missing value in  $Gaia\_G\_Magnitude$.
MICE is used to fill the missing values of these AGNs.
In Fig. \ref{fig:mice_imputations} we show the distributions of 
$Log\nu$, $Log\nu f\nu$, and $Gaia\_G\_Magnitude$ with and without MICE. 
The quality of the MICE imputations can be evaluated in part by comparing the original distribution of a variable and its distribution with imputations.
If the imputations alter the distribution, the results cannot be trusted and would require additional precautions or measures to deal with the missing values. 
However, as can be discerned from the plots (Fig. \ref{fig:mice_imputations}), the MICE imputations are indeed following the underlying distribution for the three predictors, and hence we confidently incorporate them into our analysis. 
We impute 465 data points, 24\% of our data set,
resulting in a training sample of 1444 AGNs and a generalization sample of 453 AGNs.
The two sets are detailed in Table. \ref{tab:sets}.

\begin{table}[H]
    \centering
    \begin{tabular}{c|c|c|c|c|c}
        Type & Training set & Generalization set & Redshift median & Redshift minimum & Redshift maximum\\
        \hline
        BLLs & 721 & 450 & 0.336 & $3.7\times10^{-5}$ & 2.82 \\
        FSRQ & 723 & 3 & 1.12 & 0.097 & 4.313 \\
        \hline
        Total & 1444 & 453 & 0.628 & $3.7\times10^{-5}$ & 4.313 \\
    \end{tabular}
    \caption{Composition of the training and generalization sets, and Redshift properties on the training set.}
    \label{tab:sets}
\end{table}

\subsection{MICE reliability analysis}
\begin{figure}[H]
    \centering
    \includegraphics[width=0.85\textwidth]{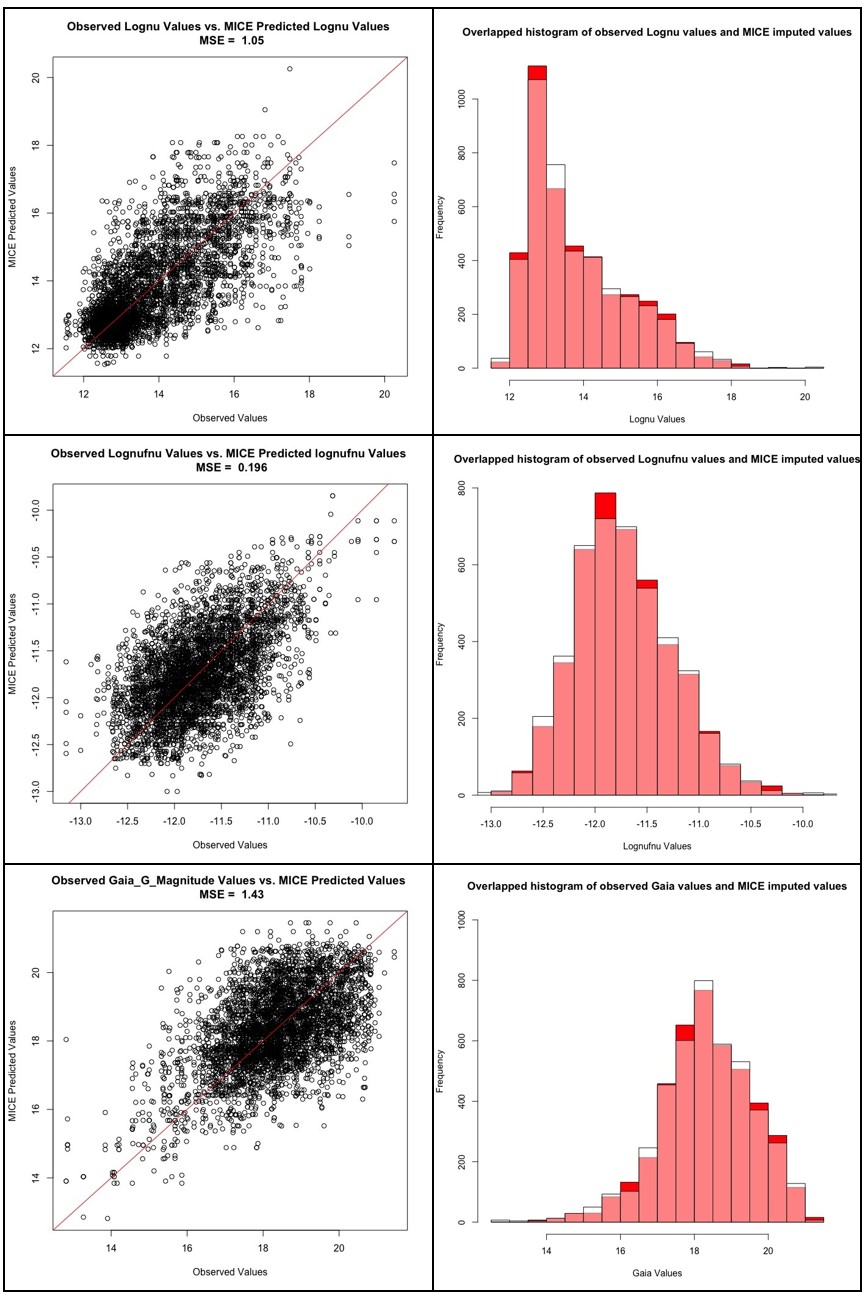}
    \caption{
    First row: The scatter plot between the observed and predicted MICE values for the $Log\nu$ predictor, followed by the overlapped histogram distributions of the same. 
    Second row: The scatter plot between the observed and predicted MICE values for the $Log\nu f\nu$ predictor, followed by the overlapped histogram distributions of the same. 
    Third row: The scatter plot between the observed and predicted MICE values for the $Gaia\_G\_Magnitude$ predictor, followed by the overlapped histogram distributions of the same. }
    \label{fig:mice_verification}
\end{figure}

\textbf{In the work by \cite{luken2021missing}, they present an extensive analysis of the reliability of MICE imputations. 
However, since they use a different dataset than ours, a similar investigation regarding the performance of MICE is essential.
Thus, we take 1432 AGNs from our catalog with no missing entries and randomly dropped 20\% of the entries from each of the three predictors which have missing entries, namely: $Log\nu$, $Log\nu f\nu$, and $Gaia\_G\_Magnitude$. 
We then impute these dropped entries using MICE, as described in Sec. \ref{sec:MICE}.
This process is repeated fifteen times, and each time a different set of random entries are dropped.}
\textbf{Furthermore, as we can see in Fig. \ref{fig:mice_verification}, the observed vs. predicted values for $Log\nu$, $Log\nu f\nu$ and $Gaia\_G\_Magnitude$ are concentrated about the $y = x$ line, with little variance. 
The mean squared error (MSE), defined as the , of the observed values vs. the MICE imputed values for $Log\nu$, $Log\nu f\nu$, and $Gaia\_G\_Magnitude$ were 1.05, 0.196, and 1.43, respectively. 
Thus, the MSEs are all small, which provides evidence that the MICE imputed effectively.}
\textbf{Note that if MICE imputes effectively, then the imputed values and observed values should come from the same distribution for each of the three variables. 
To check this, we performed a Kolmogorov-Smirnov (KS) test on the observed vs. MICE imputed values for each of the three variables with missing entries. 
The p-values of the KS test for $Log\nu$, $Log\nu f\nu$, and $Gaia\_G\_Magnitude$ were 0.744, 0.5815, and 0.6539, respectively. 
Since each of these p-values is above 0.05, we cannot reject the null-hypothesis; namely, we conclude that the observed values and the MICE imputed values come from same distributions for any of the three variables. 
As shown in Fig. \ref{fig:mice_verification}, the overlapped histogram of the observed vs. MICE imputed values for $Log\nu$, $Log\nu f\nu$ and $Gaia\_G\_Magnitude$ are each very similar, which reinforces the findings of the KS test - namely, that they are from the same distribution.
This provides additional proof for the accuracy, and reliability of the MICE imputations.}

\subsection{\textbf{with O1 Variables}}\label{sec:O1_w_MICE}
The O1 variable set consists of 12 predictors, including the categorical variable $LabelNo$, which distinguishes between BLLs and FSRQs.
LASSO chooses the best predictors from within this set for each fold in the 10fCV as explained in Sec. \ref{sec:featureEngineering}.

\begin{figure}[H]
    \centering
    \includegraphics[width=\textwidth]{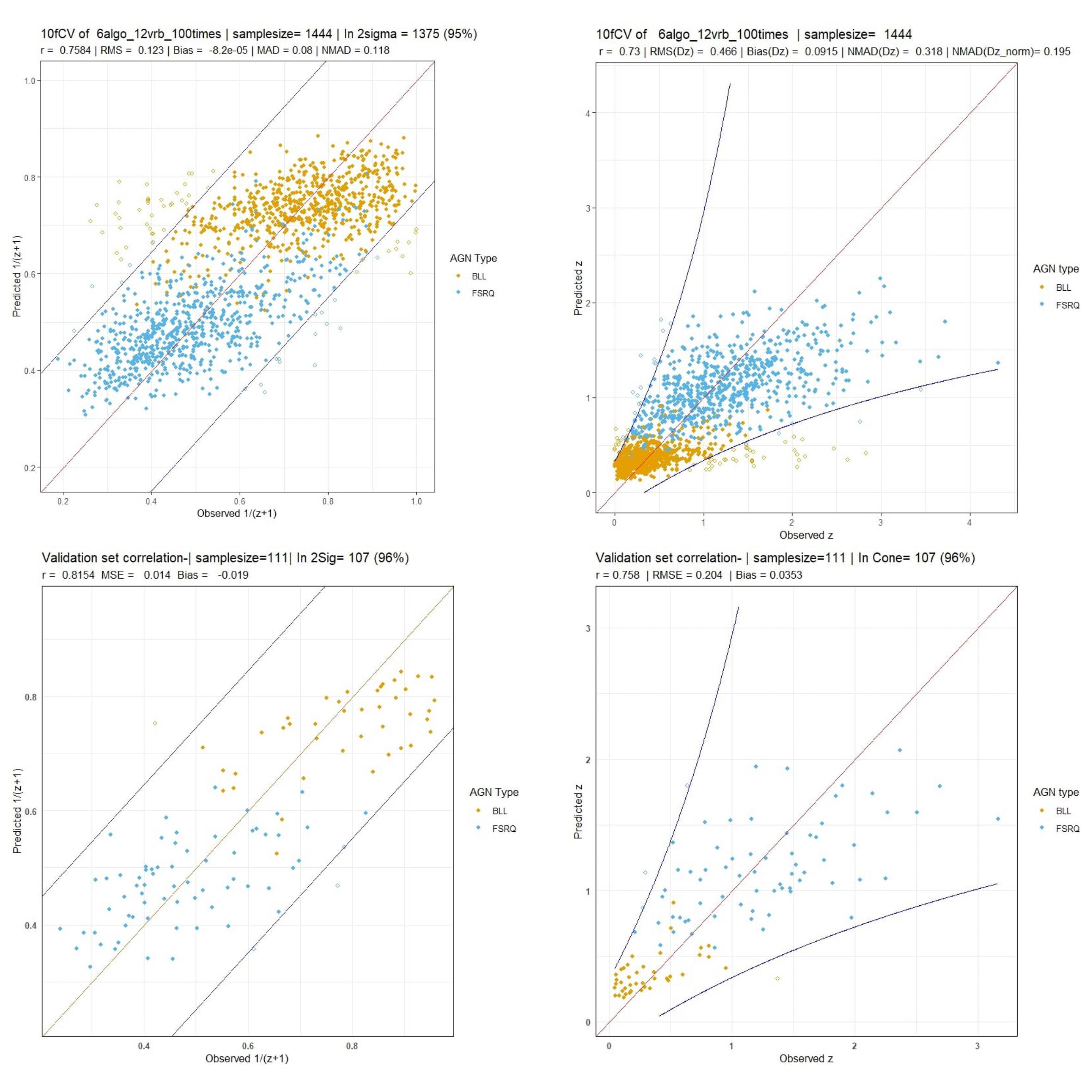}
    \caption{These plots are for the O1 predictors case.
    Top left and right panels: 
    The correlation plots between the observed and predicted redshift from 10fCV in the $\frac{1}{z+1}$ and linear scales, respectively.
    Bottom left and right panels:
    The validation set correlation plots in the $\frac{1}{z+1}$ and linear scales, respectively.
    }
    \label{fig:O1_w_MICE_1}
\end{figure}

\begin{figure}[H]
    \centering
    \includegraphics[width=\textwidth]{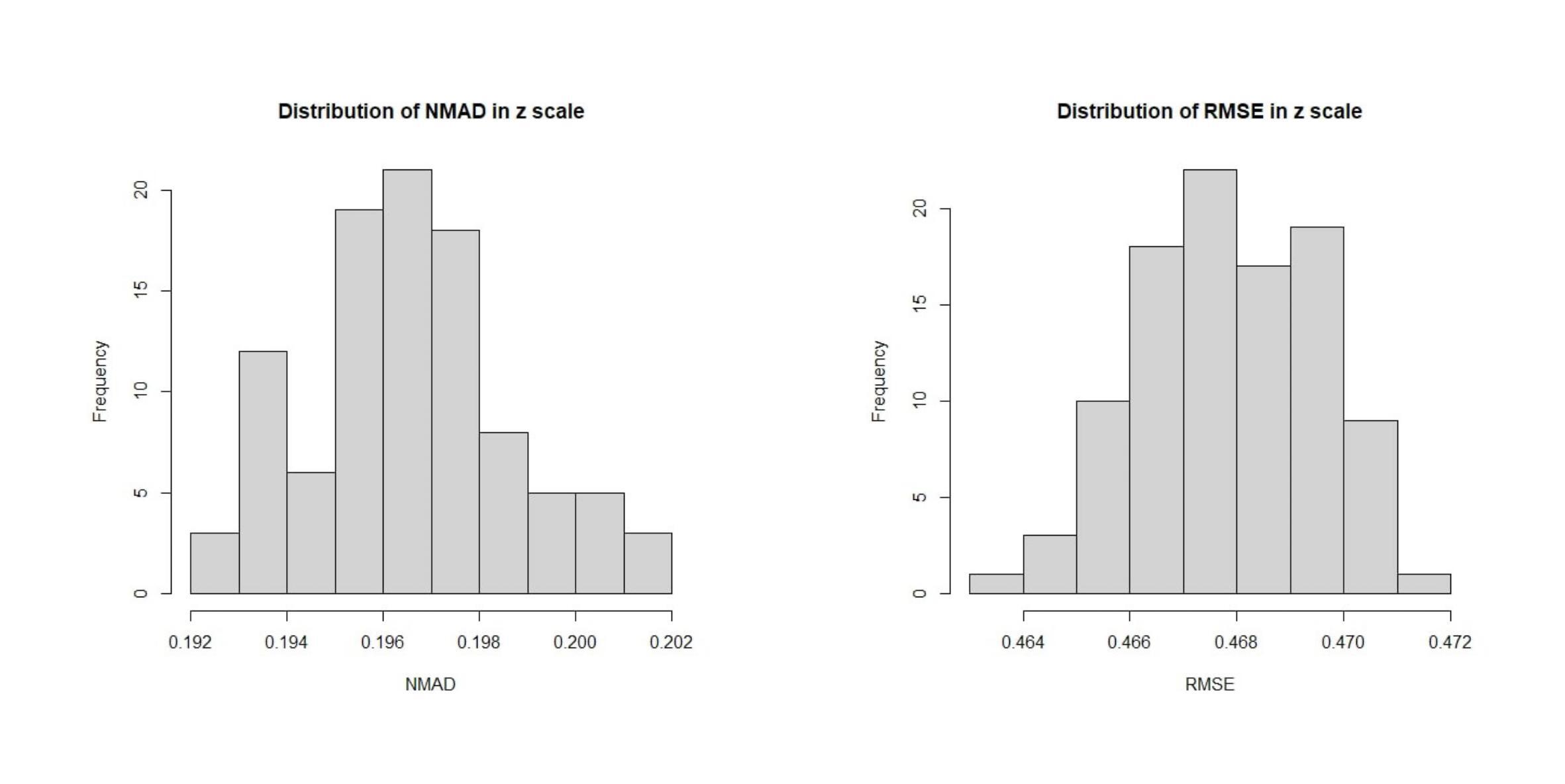}
    \caption{Here we present the distribution across one hundred iterations of 10fCV of the $\sigma_{NMAD}$ and RMSE for the O1 case.
    Left Panel: Distribution of $\sigma_{NMAD}$ in linear scale. Note that $\sigma_{NMAD}$ is denoted as NMAD in the plots.
    Right Panel: Distribution of RMSE in linear scale.
    }
    \label{fig:NMAD_RMSE_O1}
\end{figure}

Using this feature set with the six algorithms mentioned, we obtain a correlation in the $1/(z+1)$ scale of 75.8\%, a $\sigma$ of 0.123, an RMSE of 0.123, and a $\sigma_{NMAD}$ of 0.118.
In the linear redshift scale (z scale), we obtain a correlation of 73\%, an RMSE of 0.466, a $\sigma$ of 0.458, a bias of 0.092, and a $\sigma_{NMAD}$ of 0.318. 
In the normalized scale ($\Delta z_{norm}$), the RMSE obtained is 0.209, bias is $6\times10^{-3}$, and $\sigma_{NMAD}$ is equal to 0.195.
The correlation plots are shown in Fig. \ref{fig:O1_w_MICE_1}, with the left panel showing the correlation in the $1/(z+1)$ scale and the right panel showing the correlation in the $z$ scale. 
We obtain a low 5\% catastrophic outlier percentage in this scenario. 
The lines in blue depict the 2$\sigma$ curves for each plot, where the $\sigma$ is calculated in the $1/(z+1)$ scale.
\newline\indent
In Fig. \ref{fig:NMAD_RMSE_O1}, we present the distributions of $\sigma_{NMAD}$ and RMSE across the one hundred iterations. 
Note that $\sigma_{NMAD}$ is written as NMAD in the plots for brevity.

\begin{figure}[H]
    \centering
    \includegraphics[width=\textwidth]{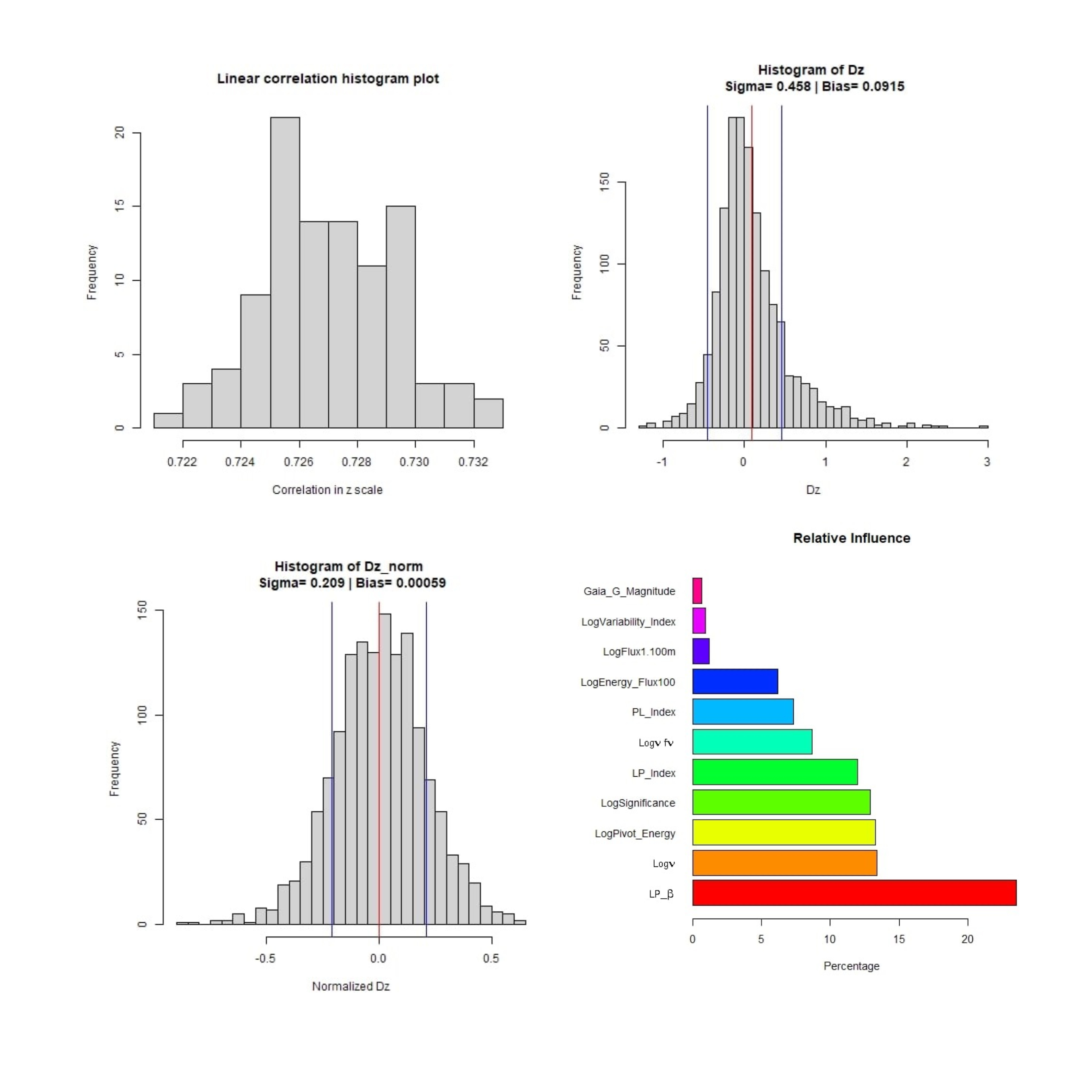}
    \caption{These plots are for the O1 predictor case.
    Top left panel: Distribution of the correlations in linear scale from the 100 iterations of 10fCV.
    Top right panel: Distribution of $\Delta z$ (Dz in the plots) with average bias (red) and sigma lines (blue).
    Bottom left panel: Distribution of the $\Delta z_{norm}$ (Normalized Dz in the plots) with the average bias (red) and sigma values (blue).
    Bottom right panel: Relative influence of the O1 predictors. The suffix of $Sqr$ implies the square of the respective predictor.
    }
    \label{fig:O1_w_MICE_2}
\end{figure}

In Fig. \ref{fig:O1_w_MICE_2}, we present the distributions of various parameters and the normalized relative influence plot of the 11 predictors - $LabelNo$ is excluded, as its a categorical variable.
The top left panel shows the variation in the linear correlation obtained from the one hundred iterations. 
The top right panel shows the distribution of $\Delta z$ along with the $\sigma$ (blue vertical line) and bias (red vertical line) values.
The bottom left panel shows the distribution of the $\Delta z_{norm}$ along with the bias and $\sigma$ presented similarly.
Finally, the barplot in the bottom right panel shows the relative influence of the 11 predictors used. $LP\_\beta$ has the highest influence, followed by $Log\nu$, $LogPivot\_Energy$, and $LogSignificance$. 
Surprisingly, $Gaia\_G\_Magnitude$ has the least influence at $\approx 1\%$, in contrast to \cite{dainotti2021predicting}, where we found it to be quite significant at $\approx 11\%$ influence.
The difference we obtain from this analysis and the previous one of \cite{dainotti2021predicting} lies in the data set and that MICE had not been used.


\subsection{\textbf{with O2 variables}}\label{sec:O2_w_MICE}
\begin{figure}[H]
    \centering
    \includegraphics[width=\textwidth]{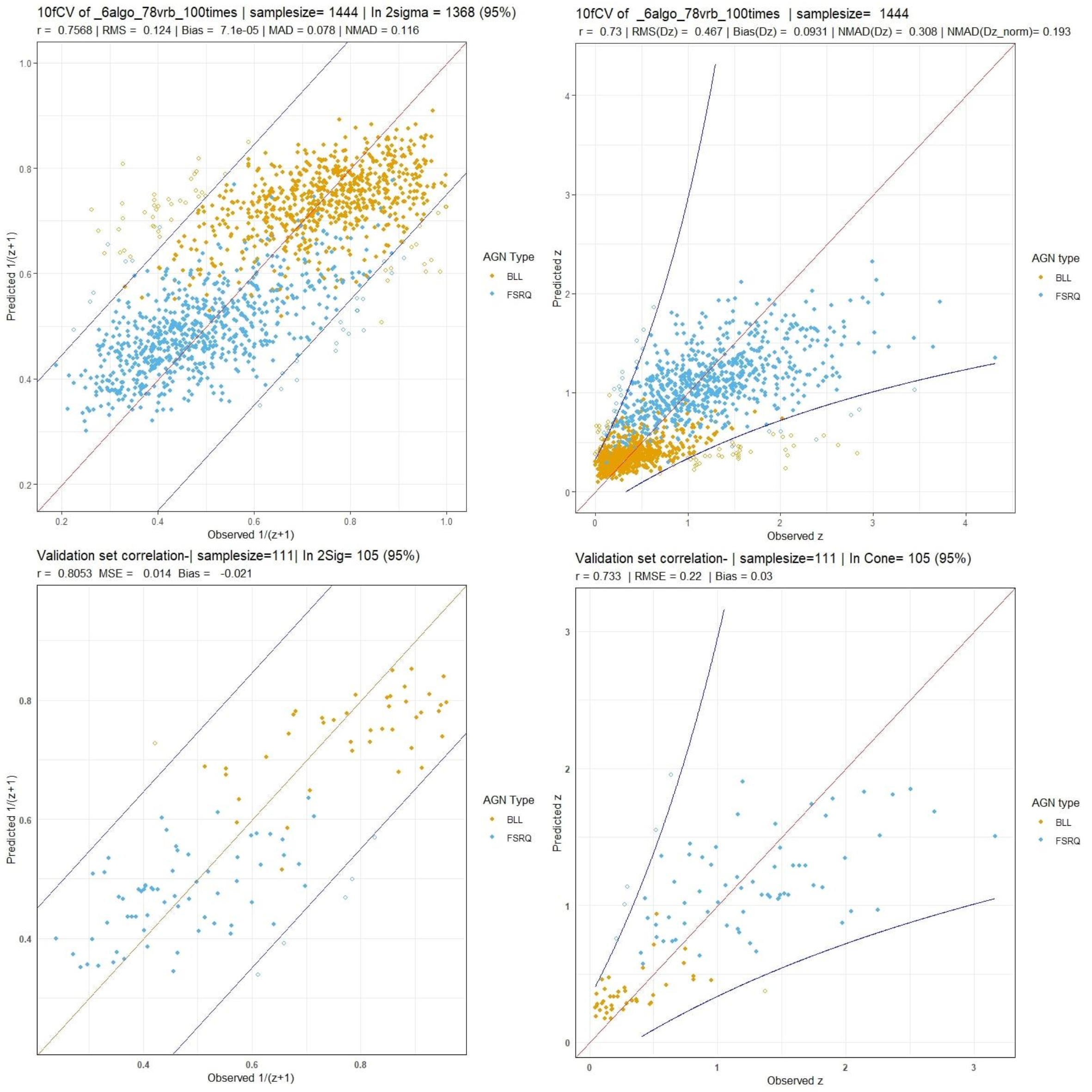}
    
    \caption{These plots are for the O2 predictor case.
    Top left and right panels: Correlation plots between observed vs. cross-validated redshift in the $\frac{1}{z+1}$ and linear scale, respectively.
    Bottom left and right panels: The validation set correlation plots in the $\frac{1}{z+1}$ and linear scale, respectively}.
    
    \label{fig:O2_w_MICE_1}
\end{figure}

The O2 variables, 78 in total, are made from cross-products of the O1 variables.
As in the O1 case, LASSO feature selection is performed on a fold-by-fold basis, 
after which the SuperLearner ensemble with the six algorithms previously mentioned is trained and makes predictions.
The cross-validation and validation correlation plots are presented in Fig. \ref{fig:O2_w_MICE_1}.

\begin{figure}[H]
    \centering
    \includegraphics[width=\textwidth]{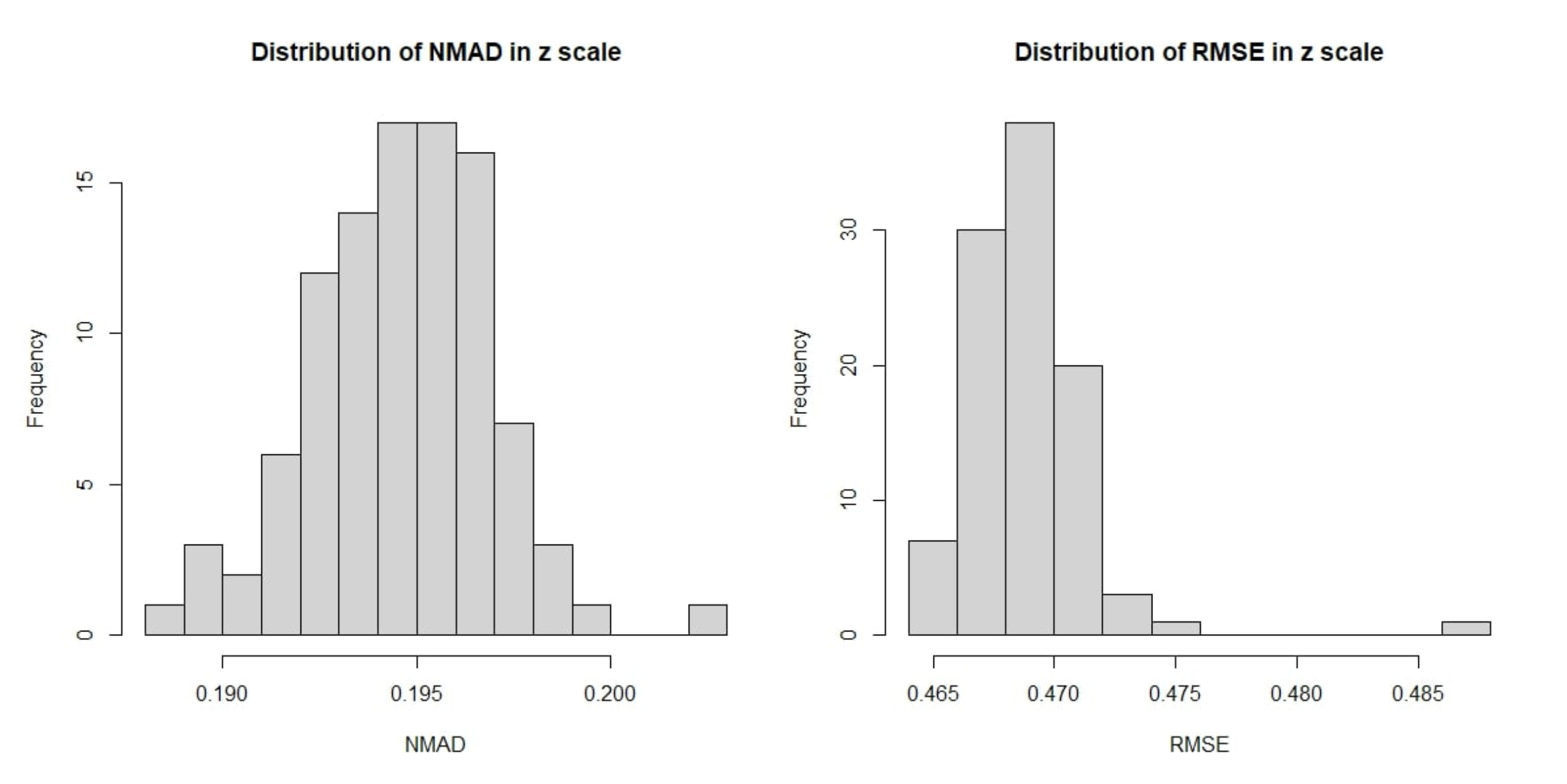}
    \caption{Here we present the distribution of the $\sigma_{NMAD}$ and RMSE for the O2 case.
    Left Panel: Distribution of $\sigma_{NMAD}$ in linear scale. Note that $\sigma_{NMAD}$ is denoted as NMAD in the plots.
    Right Panel: Distribution of RMSE in linear scale.
    }
    \label{fig:NMAD_RMSE_O2}
\end{figure}

As shown in the previous section, we have correlation plots in the $1/(z+1)$ scale and the $z$ scale. 
In the $1/(z+1)$ scale, we get a correlation of 75.6\%, RMSE of 0.124, and $\sigma_{NMAD}$ of 0.116.
In the $z$ scale, we obtain a correlation of 73\%, RMSE of 0.467, and $\sigma_{NMAD}$ of 0.308.
We obtain the statistical parameters for $\Delta z$: an RMSE of 0.467, a $\sigma$ of 0.458, a bias of 0.093, and a $\sigma_{NMAD}$ of 0.308.
For $\Delta z_{norm}$, we obtain an RMSE of 0.21, a bias of $7\times10^{-4}$, and a $\sigma_{NMAD}$ of 0.193.
We have a similar catastrophic outlier percentage (5\%) as the O1 variable case, although the number of AGNs predicted outside the 2$\sigma$ cone is seven AGNs more.
This discrepancy can be attributed to the randomness inherent in our calculations and additional noise introduced by the O2 predictors.
\begin{figure}[H]
    \centering
    \includegraphics[width=\textwidth]{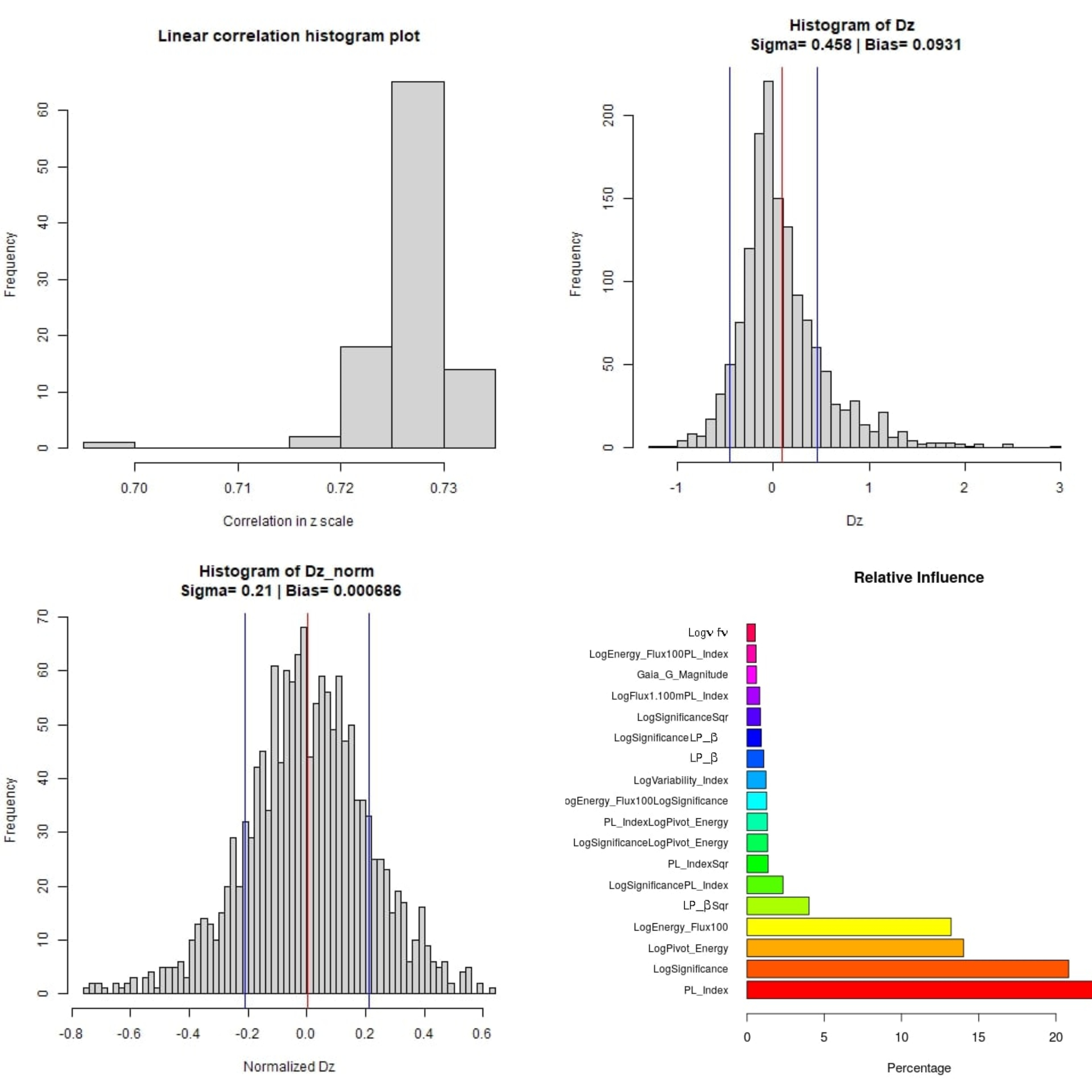}
    \caption{These plots are for the O2 predictor case.
    Top left panel: Distribution of the correlations in linear scale from the 100 iterations of 10fCV.
    Top right panel: Distribution of $\Delta z$ (Dz in the plots) with average bias (red) and sigma lines (blue).
    Bottom left panel: Distribution of the $\Delta z_{norm}$ (Normalized Dz in the plots) with the average bias (red) and sigma values (blue).
    Bottom right panel: Relative influence of the O2 predictors, above cutoff of 0.1.
    }
    \label{fig:O2_w_MICE_2}
\end{figure}

In Fig. \ref{fig:NMAD_RMSE_O2}, we show the distributions of $\sigma_{NMAD}$ and RMSE. Note that there is an outlier during the analysis, which leads to the unusually high RMSE value seen in the distribution.
\newline\indent
Figure. \ref{fig:O2_w_MICE_2} shows the distribution plots for various parameters. The top left panel shows the distribution of the correlations across the one hundred iterations. 
There is an outlier in the distribution of the correlation plot, corresponding to the distributions of RMSE in Fig. \ref{fig:NMAD_RMSE_O2}.
This scenario only happens with the O2 variable set and with MICE imputations. 
Apart from this fluctuation, most of the correlations lie around 73\%.
The histogram distribution plots for $\Delta z$ (top right) and $\Delta z_{norm}$ (bottom left) show a similar spread as in the case of the O1 variable set. 
We only present predictors with influence greater than 0.5\% in the relative influence plot.
In this case, $PL\_Index$ turns out to have the highest influence, over 20\%, followed by $LogSignificance$, $LogPivot\_Energy$, and $LogEnergy\_Flux100$.
\newline\indent
\newline\indent

We note that out of the 11 O1 predictors with relative influences, only 3 have less than 5\% influence, and out of the 78 O2 predictors, only 4 have greater than 5\% influence. 
Thus, the majority of the O2 predictors do not seem to provide much additional information about the redshift.


\textbf{
In Tables \ref{tab:comparision1} and \ref{tab:comparision2} we provide a comparision between the results obtained in the two experiments we have here with MICE, and one without MICE imputations. The latter results have been taken from \cite{narendra2022predicting}. 
}

\begin{table}[]
    \centering
    \begin{tabular}{c|c|c|c}
        Metric & SL with O1 & SL with O2 & Without MICE\\
        \hline
        \hline
        $r$ & 0.758 & 0.757 & 0.781 \\
        \hline
        RMSE & 0.123 & 0.124 & 0.119\\
        \hline
        Bias & $-8.2\times10^{-5}$ & $7.1\times10^{-5}$ & $4\times10^{-4}$ \\
        \hline
        $\sigma_{NMAD}$ & 0.118 & 0.116 & 0.113\\
        \hline
        $\sigma$ & 0.209 & 0.210 & 0.119 \\
    \end{tabular}
    \caption{Comparision of the statistical metrics across the different experiments performed. These have been calculated for the $1/(z+1)$ scale.}
    \label{tab:comparision1}
\end{table}

\begin{table}[]
    \centering
    \begin{tabular}{c|c|c|c}
        Metric & SL with O1 & SL with O2 & Without MICE \\
        \hline
        \hline
        $r$ & 0.73 & 0.73 & 0.74\\
        \hline
        RMSE ($\Delta z$) & 0.466 & 0.467 & 0.467 \\
        \hline
        Bias ($\Delta z$) & 0.0915 & 0.0931 & 0.095\\
        \hline
        $\sigma_{NMAD}$ ($\Delta z$) & 0.318 & 0.308 & 0.321\\
        \hline
        $\sigma$ ($\Delta z$) & 0.458 & 0.458 & 0.458\\
        \hline
        Bias ($\Delta z_{norm}$) & $5.9\times10^{-4}$ & $6.9\times10^{-4}$ & $9.6\times10^{-4}$ \\
        \hline
        $\sigma_{NMAD}$ ($\Delta z_{norm}$) & 0.195 & 0.193 & 0.195 \\
        \hline
        $\sigma$ ($\Delta z_{norm}$) & 0.209 & 0.210 & 0.208
    \end{tabular}
    \caption{Comparision of the statistical metrics across the different experiments performed. These have been calculated for the $z$ scale.}
    \label{tab:comparision2}
\end{table}

\section{\textbf{Discussions and Conclusions}}\label{sec:D&C}


In \cite{dainotti2021predicting}, the correlation between the observed and predicted redshift achieved with a training set of 730 AGNs was 71\%, with RMSE of 0.434, $\sigma_{NMAD}$($\Delta z_{norm}$) of 0.192, and a catastrophic outlier of 5\%.
Here, with the use of an updated 4LAC catalog, O1 predictors, and the MICE imputation technique, along with additional ML algorithms in the SuperLearner ensemble, we achieve a correlation of 73\% between the observed and predicted redshift, an RMSE of 0.466, $\sigma_{NMAD}$($\Delta z_{norm}$) of 0.195 and a catastrophic outlier of 5\%.
Although the RMSE and $\sigma_{NMAD}$($\Delta z_{norm}$) are increasing by 7\% and 1.5\%, respectively, we are able to maintain the catastrophic outliers at 5\%, while increasing the correlation by 3\%. These results are achieved with a data sample 98\% larger than the one used by \cite{dainotti2021predicting}.
Note that this achievement is not trivial, as a larger data set does not guarantee favourable results.
\newline\indent
With the O2 predictor set, we obtain a similar correlation of 73\% between the predicted and observed redshifts. However, compared to the O1 case, the RMSE goes up by 0.2\% to 0.467 and the $\sigma_{NMAD}$($\Delta z_{norm}$) goes down by 1\% to 0.193. The catastrophic outlier percentage is maintained at 5\% in both cases.


The most influential O1 predictors in this study were $LP\_\beta$, $Log\nu$, $LogPivot\_Energy$, $LogSignificance$, $LP\_Index$, $PL\_Index$, and $LogEnergy\_Flux$, each of which has a relative influence greater than $5\%$. 
$LP\_\beta$ was also the most influential predictor in \cite{dainotti2021predicting}, followed by 
$LogPivot\_Energy$, $LogSignificance$, $LogEnergy\_Flux$, and $Log\nu$. The main difference in the relative influences of the predictors in these studies is that in the O1 case with MICE, $LP\_Index$ and $PL\_Index$ are the 5th and 7th most influential predictors, respectively, while in \cite{dainotti2021predicting}, they were not influential.
\newline\indent
Among the O2 predictors, $PL\_Index$ is the most influential, followed by $LogSignificance$, $LogPivot\_Energy$, and $LogEnergy\_Flux$, each of which has a relative influence greater than $5\%$. Note that the only O2 predictors with influence greater than $5\%$ are those we have just listed and they are also O1 predictors. 
\textbf{When additional variables are added it is not guaranteed that the most influential variables will be kept the same.  This is true for both parametric and non-parametric models.  The influence is a measure of how much your improvement in the prediction changes when you remove one variable in relation to the presence of the other variables.  Thus,  these measures  depend  on  the  other  variables  in  the  model  and  are different when O2 variables are added.}
We can conclude from these results that the O1 predictors contain most of the predictive information for redshift, in the case of the 4LAC catalog.
Furthermore, we note that obtaining results with the O2 set takes more time than with the O1 set due to the larger list of predictors.
However, in other catalogs, such O2 predictors might perform better and be an avenue worth exploring in the future.
\newline\indent
\textbf{Here, we use MICE on the O1 variables, because this allows MICE to act on three variables which present missing entries. In this way, we can control the effectiveness of MICE and the results. We agree with the referee that imputing the MICE in the cross products would imply an imputation on variables that are currently not defined and most importantly would allow more uncertainty when the cross products would involve for example two variables with missing entries. 
If we had used MICE in the O2 parameters we would have had a large number of imputation which would be less controllable.}
From these results, we can discern that the MICE imputation technique is a robust method to mitigate the issue of missing entries in a catalog while maintaining the predictive power of the data.

\section*{Data and code availability}
We intend to provide the code to anyone who is interested in pursuing this avenue of research upon reasonable request to the corresponding author. The data is available on the Second Fermi AGN Catalog, see \url{https://fermi.gsfc.nasa.gov/ssc/data/access/lat/10yr_catalog/}.

\acknowledgments
This work presents results from the European Space Agency (ESA) space mission, Gaia. Gaia data are being processed by the Gaia Data Processing and Analysis Consortium (DPAC). Funding for the DPAC is provided by national institutions, in particular the institutions participating in the Gaia MultiLateral Agreement (MLA). The Gaia mission website is \url{https://www.cosmos.esa.int/gaia}. The Gaia archive website is \url{https://archives.esac.esa.int/gaia}.
M.G.D. thanks Trevor Hastie for the interesting discussion on overfitting problems.
We also thank Raymond Wayne for the initial computation and discussions about balanced sampling techniques which will be implemented in subsequent papers.
We also thank Shubbham Bharadwaj for helping create four of the plots with correct fonts and symbols.
This research was supported by the Polish National Science Centre grant UMO-2018/30/M/ST9/00757 and by Polish Ministry of Science and Higher Education grant DIR/WK/2018/12.
This research was also supported by the MNS2021 grant N17/MNS/000057 by the Faculty of Physics, Astronomy and Applied Computer Science, Jagiellonian University


\bibliographystyle{Frontiers-Harvard}
\bibliography{refs}
\end{document}